\documentclass[iop]{emulateapj-rtx4}

\shorttitle{Energy-dependent eclipsing in IC10 X-1}
\shortauthors{Barnard et al.}

\begin{document}


\title{Energy-dependent evolution in IC10 X-1: hard evidence for an extended corona, and implications}


\author{R. Barnard and J.~F. Steiner and A.~F.  Prestwich}
\affil{Harvard-Smithsonian Center for Astrophysics (CFA), Cambridge MA 02138}
\and
\author{I.~R. Stevens}
\affil{School of Physics and Astronomy, University of Birmingham, Birmingham B15 2TT, UK}
\and
\author{J.~S. Clark and U.~C. Kolb}
\affil{The Open University, Milton Keynes, UK}


\begin{abstract}
We have analyzed a $\sim$130 ks XMM-Newton observation of the dynamically confirmed black hole + Wolf-Rayet (BH+WR) X-ray  binary (XB) IC10 X-1, covering $\sim$1 orbital cycle.  This system experiences periodic intensity dips every $\sim$35 hours. We find that energy-independent evolution is rejected at a $>$5$\sigma$ level.
The spectral and timing evolution of IC10 X-1 are best explained by a compact disk blackbody and an extended Comptonized component, where the thermal component is completely absorbed and the Comptonized component is partially covered during the dip.  
We consider three possibilities for the absorber: cold material in the outer accretion disk, as is well documented for Galactic neutron star (NS) XBs at high inclination; a stream of stellar wind that is enhanced by traveling through the L1 point; and a spherical wind. 
We estimated the corona radius ($r_{\rm ADC}$) for IC10 X-1 from the dip ingress to be $\sim$10$^6$ km, assuming absorption from the outer disk,  and found it to be consistent with the relation between $r_{\rm ADC}$ and 1--30 keV luminosity observed in Galactic NS XBs that spans 2 orders of magnitude.   For the other two scenarios, the corona would be larger. Prior BH mass ($M_{\rm BH}$) estimates range over 23--38 $M_\odot$, depending on the inclination and  WR mass. For disk absorption, the inclination, $i$, is likely to be $\sim$60--80$^{\circ}$, with $M_{\rm BH}$  $\sim$24--41 $M_{\odot}$.  Alternatively, the L1-enhanced wind requires $i$$\sim$80$^{\circ}$, suggesting $\sim$24--33 $M_\odot$. For a spherical absorber, $i$$\sim$40$^{\circ}$, and $M_{\rm BH}$ $\sim$50--65 $M_\odot$. 
\end{abstract}





\section{Introduction}

IC10 X-1 is a dynamically confirmed black hole (BH) + Wolf-Rayet (WR) binary, which exhibits $\sim$35 hour periodic modulation in the X-ray and optical lightcurves \citep{prestwich07,silverman08}; the likely association between the BH and WR was first reported by \citet{clark04}. Its mass function is 7.6$\pm$1.3 $M_\odot$, giving a BH mass $\sim$23--38 $M_\odot$ for the likely range in WR mass and system inclination  \citep{silverman08}.

IC10 X-1 was previously observed by XMM-Newton in 2003. \citet{wang05} reported large intensity variation over $\sim$10 ks: a factor of up to $\sim$6 in the 0.5--2.0 keV band, and a factor 4 in the 2.0--7.5 keV band; they suggested that this could be due to an eclipse by the companion star. They found that the emission spectrum was well characterised by a multi-temperature disk blackbody, and rejected a simple power law model; the observed 1.18$\pm$0.02 keV inner disk temperature was consistent with a black hole binary in its thermal dominant state \citep{remillard06}. However, this fit required a metalicity $<$0.01 Solar because Solar abundance models failed to provide acceptable fits. They inferred a BH mass of $\sim$4 $M_{\odot}$ for a non-spinning BH, and a mass $\sim$6 times greater for a rapidly spinning BH. \citet{wang05} attempted to compare the spectra in and out of the eclipse, but there were too few counts to observe any significant differences.

IC10 X-1 was also observed several times with Chandra and Swift, leading to the discovery of its orbital period \citep{prestwich07}.  The 0.3--10 keV lightcurve of IC10 X-1 exhibits deep, periodic intensity dips that have been interpreted as eclipses by the WR; however, the full dip profile had never previously  been seen.

IC10 X-1 was observed with XMM-Newton for $\sim$130 ks, equivalent to $\sim$1 orbital cycle, in 2012 June (ObsID 0693390101003, PI T. Strohmayer). \citet{strohmayer13} found the intensity dip to be asymmetric, with ingress and egress times of $\sim$3.3 and 4.6 hours, with a maximum eclipse lasting $\sim$5.2 hours. Additionally, they found $\sim$7 mHz quasiperiodic oscillations \citep[see also][]{pasham13}. 

Such evolution, with the egress substantially longer than the ingress, is highly reminiscent of the behavior observed in the high inclination Galactic neutron star binaries known as the dipping sources \citep[see e.g.][]{parmar86}. It is possible that IC10 X-1 experiences similar processes, and we therefore present a brief review of the Galactic dipping sources. When discussing these systems, the term ``eclipse'' is reserved for energy-independent occultation by the companion star; the energy-dependent photo-electric absorption by other material is referred to as ``dipping'', and individual events are called ``dips''. However we note that HMXBs such as Vela X-1 can also exhibit such asymmetric lightcurves due to asymmetries in the wind.

It is  of fundamental importance  to determine whether this intensity modulation is energy-dependent or not; energy-dependent dipping cannot be due to occultation of the X-ray source by the star itself.

\subsection{Dipping NS XBs in our Galaxy}

Around 10 Galactic low mass X-ray binaries (LMXBs) exhibit periodic intensity dips in their X-ray lightcurves on the orbital period. This behavior is caused by photoelectric absorption of X-rays by cold material in the outer accretion disk, such as the bulge where the accretiion stream collides with the disk \citep{white82}. The spectral evolution during dipping is well understood as the progressive covering of a point-like thermal component (from the NS) and an extended Comptonized component by an extended absorber \citep[see e.g][]{church95,barnard01,smale01}. The thermal component sees little additional absorption until the covering fraction of the Comptonized component reaches $\sim$50\%; then the absorption of the thermal component jumps to very large values \citep[see e.g.][]{barnard01,smale01,smale02}. We note that some dipping sources exhibit shallow ``shoulders'' on either side of the dip; these are energy-independent and consistent with electron scattering by the ionized outer region of the absorber \citep[e.g.][]{smale01,smale02}.

4U1624$-$49 is a Galactic dipping source with orbital period $\sim$21 hours that does not exhibit eclipses. Its $\sim$1--10 keV lightcurve from the 1985 March 25 Exosat observation exhibited flat-bottomed dipping \citep{church95} that strongly resembles the ``eclipses'' observed in the 0.3--10 keV XMM-Newton lightcurve of IC10 X-1. However, RXTE observations of 4U 1624$-$49 exhibit no signs of dipping above 15 keV \citep{smale01}. It is therefore possible that the periodic intensity dips observed in IC10 X-1 could be caused by material in the outer accretion disk instead of the donor star. In some cases, the duration of dipping can be extremely high: the transient, eclipsing, Galactic NS XB EXO 0748$-$676 suffered dipping for $\ga$70\% of the orbital cycle during a 1993 May ASCA observation \citep{church98}.

One common property of the absorption features in dipping  sources is that the egress is shallower than the ingress (and Strohmayer et al. 2013 noted this property in the IC10 X-1 lightcurve). Since the thermal and non-thermal components of the dipping sources suffer increased absorption at different rates during dipping, it is clear that  the ingress is caused by a change in covering fraction, rather than a change in the density of the absorber. It is therefore possible to estimate the scale of the Comptonized region (i.e. the corona). \citet{church04} estimated the corona sizes for 5 dipping sources with 1--30 keV luminosities $\sim3\times 10^{36}$--$1.4\times 10^{38}$ erg s$^{-1}$ using the following approximation:
\begin{equation}
\frac{2\pi r_{\rm AD}}{P} = \frac{2r_{\rm ADC}}{\Delta t},
\end{equation}
where $P$ is the orbital period, $\Delta t$ is the ingress time, $r_{\rm AD}$ is the radius of the accretion disk, and $r_{\rm ADC}$ is the radius of the disk corona. \citet{church04} estimated $r_{\rm AD}$ to be 80\% of the Roche lobe equivalent radius, $r_{\rm L1}$, based on  values 0.74--0.84$r_{\rm L1}$ from the simulations of  \citet{armitage96} and 0.9$r_{\rm L1}$ from \citet{frank02}. The radius of a sphere with volume equivalent to the Roche lobe of the accretor, $r_{\rm L1}$, was calculated from the following expression derived by \citet{eggleton83} (accurate to 1\%):
\begin{equation}
r_{\rm L1} = \frac{0.49a\left(M_1/M_2\right)^{2/3}}{0.6\left(M_1/M_2\right)^{2/3}+\ln\left[1+\left(M_1/M_2\right)^{1/3}\right]},
\end{equation}
where $a$ is the binary separation, and $M_1$ and $M_2$ are the masses of the accretor and donor respectively. 

\citet{church04} found corona radii $\sim$20,000--700,000 km ($\sim$10--60\% of the disk, with 15\% typical); they also found that $r_{\rm ADC}$ $\propto$ $L_{1-30}^{0.88\pm0.16}$ with 99\% confidence. They presented evidence that $r_{\rm ADC}$ is comparable to the Compton radius, i.e. the maximum radius for hydrostatic equilibrium; this radius is expected to increase with luminosity because the electron temperature is reduced by the higher flux of soft photons.

More recently, the extended corona model was used to explain the odd dipping behavior exhibited by Cygnus X-2, where the blackbody component is unaffected \citep{balucinska-church11}. Cygnus X-2 was simultaneously observed in radio, optical and X-ray with XMM-Newton, Chandra, and the European VLBI Network, and a combination of CCD and grating spectroscopy was used to probe the dip evolution. The blackbody emission was unaffected by these dips, while $\sim$40\% of the Comptonized component from the extended corona suffered extra absorption equivalent to 3$\times 10^{23}$ H atom cm$^{-2}$. The dips occured at phase 0.35 in the orbital cycle, opposite the bulge where the accretion stream impacts the disk; \citet{balucinska-church11} demonstrated that such material viewed at inclinations $>$60$^\circ$ could cover large parts of the corona without reaching the NS.

 \citet{schulz09} found independent evidence for extended coronae in NS LMXBs from Chandra grating spectroscopy of Cygnus X-2. They examined several emission lines between 1.5--25\AA\, with the high energy transmission grating. These emission lines were observed at rest, and had widths between 1120 and 2730 km s$^{-1}$; these corresponded to a corona radius $\sim$20,000--110,000 km.

 One model describing such a system is that of \citet{haardt93}, where the hot corona is coplanar with the cold accretion disk; photons from the cold disk feed the corona, while high energy photons from the corona heat the disk.  
  Indeed XMM-Newton observations of several bright extra-galactic X-ray binaries have revealed spectra that strongly prefer models where the Comptonized component dominates the low-energy spectrum to models where the thermal component dominates \citep{barnard10,barnard11b,barnard13}.

We note that \citet{boirin05} suggested an alternative explanation for dipping after discovering absorption lines in XMM-Newton observations of 4U\thinspace 1323$-$62 that grow stronger during dipping. They proposed that variations in the continuum and absorption lines could be produced by using a warm, highly-ionized absorber; their model does not require partial covering, or an extend corona. \citet{boirin05} were able to get acceptable fits to individual spectra of 4U\thinspace 1323$-$62 at different stages of dipping. However,  \citet{diaztrigo06} examined XMM-Newton observations of several other Galactic dipping LMXBs, fitting different stages of dipping simultaneously; they quote best fits with  $\chi^2$/dof =1.15--1.3, but the many degrees of freedom (720--1340) meant that the fits were formally unacceptable, and it is unclear whether this is due to the model or uncertainties in calibration etc.. Furthermore, the warm absorber model cannot account for the dipping in Cygnus X-2 where the blackbody is unaffected \citep{balucinska-church11}. The discovery of a highly-ionized absorbing region in dipping XBs is very  interesting; however, the warm absorber model is not a credible replacement for partial covering. While we are unlikely to see such features in IC10 X-1 from the disk, it is possible that the warm WR  wind may act in a similar way.

\subsection{Possible scenarios for IC10 X-1}

One substantial difference between IC10 X-1 and the dipping LMXBs is that the donor is a Wolf-Rayet rather than a Roche lobe-filling low mass star. \citet{prestwich07} showed that the  WR could not be Roche lobe filling, and expected only a tenuous accretion disk from the stellar wind. Such a tenuous disk may not support the extended corona or the substantial absorber observed in the  Galactic NS dipping sources. However \citet{barnard08a} proposed that IC10 X-1 contains a substantial accretion disk that is fed by the WR wind rather than Roche lobe overflow.

The BH HMXBs Cygnus X-1, LMC X-1, and LMC X-3 have  longer orbital periods than IC10 X-1, yet  their X-ray spectra still exhibit substantial disk emission \citep[see e.g.][and references within]{remillard06}; LMC X-3 is powered by Roche lobe overflow, but Cygnus X-1 and LMC X-1 are likely powered by a wind-fed disk, and IC10 X-1 is too. 

We note that  the X-ray  emission from Galactic NS HMXB Vela X-1 is dominated by 2-3 power laws with the same photon index but with different normalizations and column densities; this is interpreted as direct emission from the neutron star as well as a scattered component from free electrons in the wind; Vela X-1 exhibits eclipses on a $\sim$9 day period, and some direct emission is still seen during eclipse \citep[see e.g.][and references within]{sako99,martinez14}.  Since IC10 X-1 has a very strong wind, it is possible that some of its emission is scattered also. Vela X-1 also exhibits strong fluorescence line emission, as well as recombination lines and continuum emisssion from highly ionzed gas \citep{sako99}, which we are unlikely to detect in IC10 X-1.

 We  consider three scenarios in this work that could produce the observed lightcurve: energy-independent occultation by the donor star, energy-dependent absorption from structure in a substantial disk, and energy-dependent absorption from the stellar wind. The latter two scenarios are indistinguishable from the X-ray data, so our spectral analysis only considers energy-independent vs. energy-dependent evolution.

\section{Observations and data analysis}
IC10 X-1 was observed by XMM-Newton on 2012 August 18 for $\sim$130 ks (Obs ID 
0693390101003, PI T. Strohmayer). We analyzed these data with the XMM-Newton Software Analysis System (SAS) version 13.0.0.  Substantial background flares occurred near the beginning and end of the observation \citep{pasham13}; we identified these events by extracting a 10--12 keV lightcurve with PATTERN==0 and FLAG==0, and highlighting  intervals with pn intensities $>$0.4 count s$^{-1}$. 

We extracted source and background spectra from the pn instrument, and analyzed these spectra with XSPEC version 12.8.1. These spectra were grouped to a minimum of 20 counts per bin. The background was accumulated from a circular region of empty sky on the same CCD as the source region.  Since the intensity dips are thought to be caused by absorption, we simultaneously fitted non-dip and dip spectra; the emission parameters were the same for the non-dip and dip spectra, but free to vary. Unfortunately, we were unable to fit the ingress or egress because the timescale of spectral evolution is significantly shorter than the exposure time required for good spectra.

We used the {\sc abund} command in XSPEC to specify a metalicity 0.15 times Solar \citep[appropriate for IC10, see][]{lequeux79,bauer04}, assuming the Solar abundances reported by \citet{lodders03}. However, XSPEC does not allow us to specify a Solar-metalicity foreground absorber when the abundance is set this way; the Galactic foreground absorption is translated into some portion of the measured low-metalicity absorption.


\begin{figure}
\epsscale{1.2}
\plotone{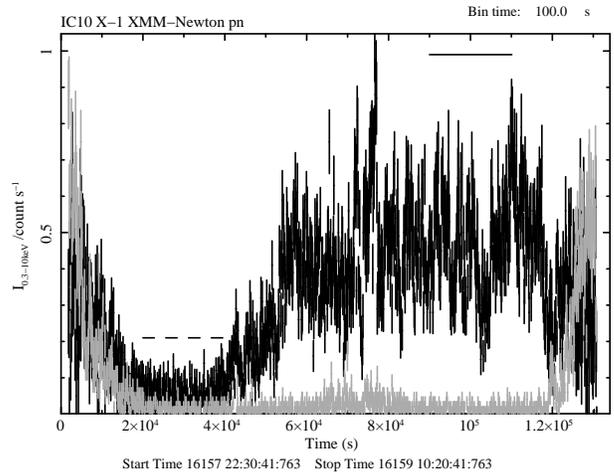}
\caption{ 0.3--10 keV background-subtracted, unfiltered pn lightcurve of IC10 X-1 (black), compared with the background lightcurve (gray).  These lightcurves have 100 s binning. The solid and dashed lines indicate the intervals included in the non-dip and dip spectra respectively.}\label{lc}
\end{figure}

\section{Results}
\label{res}

\begin{figure*}
\epsscale{1.1}
\plotone{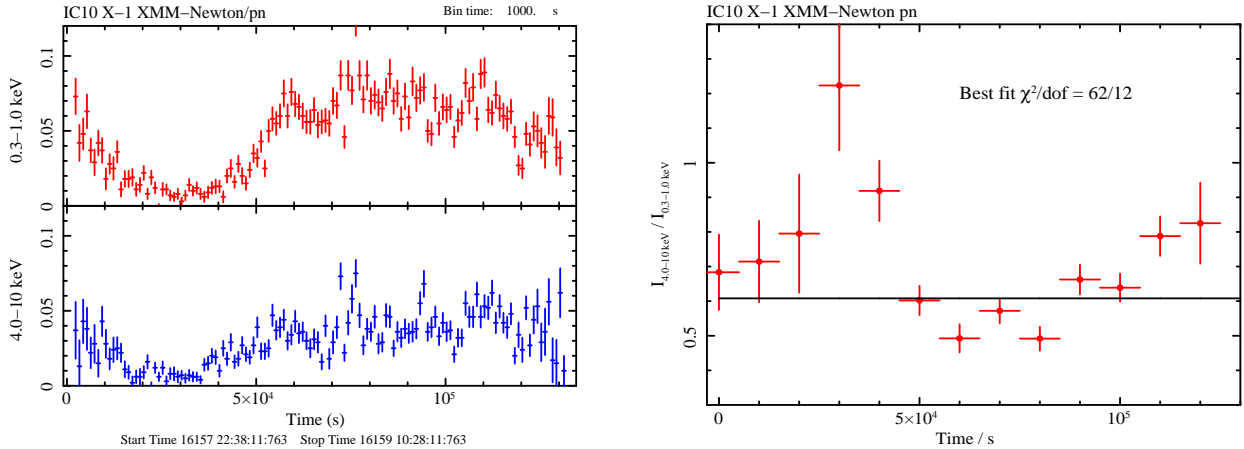}
\caption{{\em Left:~}XMM-Newton pn lightcurves of IX10 X-1 near the dipping interval in 0.3--1.0 keV and 4.0--10 keV bands. The lightcurves have 1000 s bins. The minima are similar for the two energy bands, even though the 0.3--10 keV band is brighter outside the dip; hence the dip appears to be energy-dependent. {\em Right:~} Hardness ratio vs. time binned to 10000 s; the best constant hardness fit yielded $\chi^2$/dof = 62/12, with a probability $\sim2\times 10^{-8}$.}\label{cols}
\end{figure*}

In Figure~\ref{lc} we present the  background-subtracted 0.3--10 keV pn intensity lightcurve  of IC10 X-1 (black), and the  background lightcurve in gray for comparison. Horizontal lines represent the intervals used when extracting spectra: the solid line represents the non-dip interval (90,000-110,000 s after T$_0$), 
and the dashed line represents the dip interval (20,000--40,000 s after T$_0$). The non-dip spectrum yielded 9250 net source counts, while the dip spectrum yielded 1150 net source counts.

Simultaneous fitting of the non-dip and dip spectra allowed us to test different models for the spectral evolution, which we discuss in turn. These dips could be energy-independent if caused by the donor star itself; alternatively they could be energy-dependent absorption features due to structure in the outer accretion disk  \citep[as seen in the Galactic dipping sources, e.g.][]{church95,barnard01}, or due to the dense stellar wind from the Wolf-Rayet donor.  

\subsection{Comparing the low and high energy lightcurves}
To test whether the absorption is energy-dependent,  we examined the low and high energy lightcurves.  To accomplish this, we created source and background lightcurves in the 0.3--1.0 and 4.0--10 keV bands.

The left panel of  Figure~\ref{cols} shows the background-subtracted 0.3--1.0 keV and 4.0--10 keV lightcurves from the pn, binned to 1000 s. The right panel shows the hardness ratio defined by 4.0--10 keV intensity divided by 0.3--1.0 keV.
The best fit value for a  constant hardness ratio yielded $\chi^2$/dof = 62/12. This has a good fit probability $\sim2\times 10^{-8}$, and is rejected at  a  5.6 $\sigma$ level. Therefore, the absorption cannot be simple occultation by the WR, and must come from a lower-density region such as the bulge in the outer accretion disk, or from the WR wind.
 
\subsection {Spectral analysis}

We tested a series of spectral models by simultaneously fitting non-dip and dip data (as defined above), with only the absorption free to vary. 
All of our spectral models include line-of-sight absorption, $N_{H}$, and  a partial covering component (PCF in XSPEC) that is described by the amount of absorption, $N_{\rm H}^{\rm c}$, and the fraction of the emission region covered, $f_{\rm c}$. For the non-dip spectrum, $N_{\rm H}^{\rm c}$ and $f_{\rm c}$ are frozen at 0. The models  also include a disk blackbody component (DISKBB in XSPEC), parameterized by inner disk temperature $kT_{\rm in}$, and some sort of Comptonization component (SIMPL, COMPTT, or PO--- short for power law--- in XSPEC). We use the CFLUX convolution model to estimate the 0.3--10 keV intrinsic luminosity  of each emission component; we adopt a 715 kpc distance to IC10 \citep{kim09}, which is subject to  $\pm$10 kpc random uncertainties and $\pm$60 kpc systematic uncertainties.

\
\subsubsection{Compact corona scenarios}
We first consider a compact corona scenario (Model I), where the emission consists of a disk blackbody and a Comptonized component, represented by TBABS*(CFLUX*DISKBB+CFLUX*COMPTT) in XSPEC. Many authors believe that the corona resides inside the accretion disk, and can only access the hottest photons at the inner edge of the disk \citep[see e.g.][]{roberts05,gs06}; hence, we set the seed photon temperature of the COMPTT component to the  inner disk temperature of the disk blackbody component. 

The best fit model yielded $kT_{\rm in}$ = 0.133$\pm$0.010 keV, with the COMPTT parameters unconstrained; $N_{\rm H}$ =  6.2$\pm$0.6$\times 10^{22}$ atom cm$^{-2}$, $N_{\rm H}^{\rm c}$  = 8.2$\pm 1.8\times 10^{24}$ atom cm$^{-2}$, $f_{\rm c}$ = 0.886$\pm$0.004, and $\chi^2$/dof =420/427.   Statistically, this model is a good fit to the spectrum; however, the inner disk temperature is very low, indicating a near maximum retrograde spin for a black hole with mass $\sim$30 $M_\odot$, which is possible but  unlikely. The total 0.3--10 keV luminosity for this fit is 1.8$\pm$0.7$\times 10^{39}$ erg s$^{-1}$, 47$\pm$18\% Eddington for a $\sim$30 $M_\odot$ BH.

\begin{figure*}
\epsscale{1.1}
\plotone{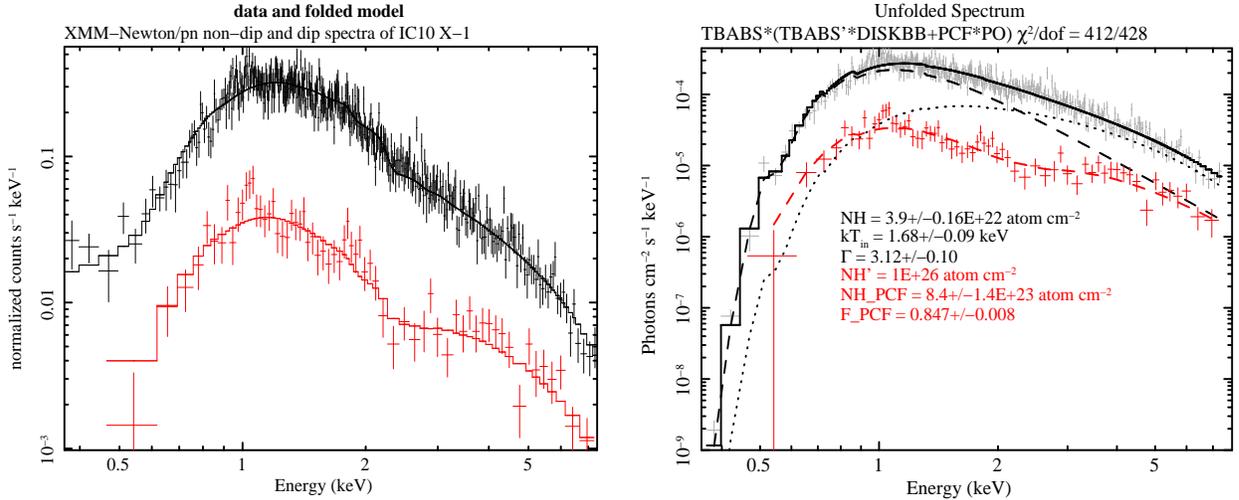}
\caption{{\em Left} Folded non-dip and dip spectra fitted with our best model (Model V). {\em Right} Unfolded non-dip and dip spectra, with our favored model: a compact thermal and an extended Comptonized component that experience different degrees of absorption during dipping. The emission spectrum is characterized by a 1.68$\pm$0.09 keV disk blackbody and a power law with $\Gamma$ = 3.12$\pm$0.10; line of sight absorption is equivalent to 3.91$\pm$0.16$\times 10^{22}$ atom cm$^{-2}$. The thermal component is removed completely from the dip spectrum while the Comptonized component is partially covered during the dip: 84.7$\pm$0.8\% of the Comptonized  emission experiences 8.4$\pm$1.4$\times 10^{23}$ atom cm$^{-2}$ extra absorption.  The thermal and Comptonized components are represented by dotted and dashed lines respectively. }\label{fold}
\end{figure*}

We next applied the SIMPL \citep{steiner09} convolution model for Comptonization (Model II), where a fraction of the thermal emission, $f_{\rm S}$ is Comptonized, and described by a power law with photon index $\Gamma$. The appropriate spectral model was TBABS*PCF*CFLUX*SIMPL*DISKBB. 

The best fit yielded  $kT_{\rm in}$ = 0.118$\pm$0.010 keV, $\Gamma$ = 2.43$\pm$0.06, $f_{\rm S}$ = 0.020$\pm$0.013;  $N_{\rm H}$ = 6.9$\pm$0.6$\times 10^{22}$ atom cm$^{-2}$, $N_{\rm H}^{\rm c}$ = 9.3$\pm$2.4$\times 10^{24}$ atom cm$^{-2}$, $f_{\rm c}$ = 0.886$\pm$0.004, and $\chi^2$/dof = 438/428. The results from this model are very similar to those obtained with the previous model, although the Comptonized component is better constrained. Again, the low temperature makes this scenario unlikely. The 0.3--10 keV luminosity is 3.7$\pm$1.7$\times 10^{39}$ erg s$^{-1}$, i.e. 100$\pm$40\% Eddington for a $\sim$30 $M_{\odot}$ BH.

We note that since SIMPL is a convolution model, we cannot treat the thermal and non-thermal components separately. We are therefore unable to apply different absorptions to the thermal and Comptonized components. This may explain why the best fit parameters resemble those for Model I, even though SIMPL does not specify a compact corona.

Finally we considered a compact corona model that includes scattering from free electrons in the wind (Model III).
 The XSPEC model used was PCF*TBABS*(CFLUX*DISKBB+CFLUX*COMPTT)+\\PCF'*TBABS'*(CFLUX*DISKBB+CFLUX*COMPTT), with the seed temperature of the COMPTT component linked with  the inner disk temperature as with Model I. The spectral shapes of the two components (i.e. inner disk temperature, electron temperature, and opacity) were the same for each component  but free to vary; the normalizations of each component were unconstrained.
 
This model fitted the spectra very well, which is perhaps unsurprising given that it contained 4 emission components  and 11 free parameters: $\chi^2$/dof = 383/422. However, the inner disk temperature (0.20$\pm$0.03 keV) indicated a strong  retrograde spin, as with previous models with a compact corona. In Table~1 we present two values each for the absorption, normalization and partial covering parameters, corresponding to the direct and scattered emission respectively. We note that for the scattered component, the thermal flux goes to zero, while the Comptonized flux is higher than for the direct component; this is probably an artifact caused by fitting the high energy excess in the dip spectrum.

\subsubsection{Extended corona models}

Since the extended corona may have  access to soft photons from the outer disk in addition to the harder photons at the inner disk \citep[e.g.][]{haardt93}, the Comptonized component is decoupled from the thermal component, and may  be represented as a power law.
 We explore two models for the extended corona scenario: one where both components suffered the same absorption, and one with different absorption for the thermal and Comptonized component.
We modeled these scenarios  as TBABS*(TBABS$'$*CFLUX*DISKBB+\\PCF*CFLUX*PO). The additional absorption suffered by the thermal component, i.e. TBABS$'$, was frozen at 0 for the non-dip spectrum.

Forcing the disk blackbody and power law components to experience the same absorption (Model IV) yielded the following results: $kT_{\rm in}$ = 1.49$\pm$0.15 keV, $\Gamma$ = 2.50$\pm$0.06; $N_{\rm H}$ = 3.38$\pm$0.16$\times 10^{22}$ atom cm$^{-2}$, $N_{\rm H}^{\rm c}$ = 640$\pm$170$\times 10^{22}$ atom cm$^{-2}$, $f_{\rm c}$ = 0.854$\pm$0.012; $\chi^2$/dof = 468/428. Such a spectrum is entirely consistent with a black hole in its steep power law state \citep{remillard06}. The 0.3--10 keV luminosity for this model is 2.6$\pm$0.2$\times 10^{38}$ erg s$^{-1}$, $\sim$7\% Eddington for a $\sim$30 $M_\odot$ BH.

Setting the in-dip blackbody absorption to 10$^{26}$ atom cm$^{-2}$ and freely fitting the power law in-dip absorption (Model V) yielded $kT_{\rm in}$ = 1.68$\pm$0.09 keV, and $\Gamma$ = 3.12$\pm$0.10; $N_{\rm H}$ = 3.91$\pm$0.15$\times 10^{22}$ atom cm$^{-2}$, $N_{\rm H}^{\rm c}$ =84$\pm$14$\times 10^{22}$ atom cm$^{2}$,  $f_{\rm c}$ = 0.846$\pm$0.008, and $\chi^2$/dof = 412/428. The 0.3--10 keV luminosity of this model is 3.7$\pm$0.3$\times 10^{38}$ erg s$^{-1}$, or $\sim$10\% Eddington for a $\sim$30 $M_\odot$ BH.

 This model yielded the best fit out of the constrained models (good fit probability 0.71) and also believable parameters for a BHC in the steep power law state \citep{remillard06}; the high disk blackbody temperature corresponds to a high positive spin, as seen in \citet{mcclintock06}. The folded and unfolded spectra are presented in the left and right panels of Fig.~\ref{fold}.

\begin{table*}
\begin{center}
\caption{Summary of spectral models I--V; for each model we give the line of sight absorption ($N_{\rm H}$); disk blackbody temperature ($kT_{\rm in}$) and normalization; photon index ($\Gamma$), normalization and Comptonization where apropriate; and $\chi^2$/dof. For Model IV we give the power law normalization of the nondip spectrum and the percentage scattered into the dip spectrum. The metalicity was set to 0.15 Solar, as appropriate for IC10 X-1 \citep{lequeux79,bauer04}. } \label{modsum}
\renewcommand{\arraystretch}{.8}
\begin{tabular}{cccccccccccc}
\tableline\tableline
Model & $N_{\rm H}$ / 10$^{22}$  & $kT_{\rm in}$ / keV & $L_{\rm DBB}^{37}$ & $\Gamma$ & $L_{\rm Comp}^{37}$ & $f_{\rm S}$ & $N_{\rm H}^{c}$ / 10$^{22}$ & $f_{\rm c}$ & $\chi^2$/dof \\
\tableline \\
Model I& 6.2(6) & 0.133(10) & 150(70) & ? & 27.5(18) & --- & 820(180) & 0.886(4) & 420/427 \\
Model II& 6.9(5) & 0.118(11) &370(170) & 2.43(6) & --- & 0.020(13) & 930(240) & 0.886(4) & 438/428 \\
Model III& 4.7(6)/19(5) & 0.20(3)& 180(90)/0(?)& ? & 8(2)/28(5)&  --- & 100(30)/$>10^4$& 0.873(8)/0.94(3) & 383/422\\
Model IV & 3.38(16) & 1.49(15) & 4.7(13) & 2.50(6)  & 21.3(16) & --- & 635(166) & 0.854(12) & 468/428\\
Model V & 3.91(16) &  1.68(9) & 8.1(4) & 3.12(10)  & 29(3) & ---  & 84(14) & 0.847(8) &  412/428 \\
\tableline
\end{tabular}

\end{center}
\end{table*}

\subsection{Summary of spectral results}
We present a summary of the best fits from Models I--V in Table~\ref{modsum}. For each model we give the line-of-sight absorption ($N_{\rm H}$), the disk blackbody temperature ($kT_{\rm in}$) and 0.3--10 keV luminosity / 10$^{37}$ erg s$^{-1}$, the photon index ($\Gamma$) and 0.3--10 keV luminosity  of the Comptonized component, degree of Comptonization where applicable ($f_{\rm S}$), plus the column density and covering fraction during the dip ($N_{\rm H}^{\rm c}$ and $f_{\rm c}$ respectively). Unconstrained parameters are indicated by '?', and parameters that do not apply to the model are indicated by '---'.  Numbers in parentheses represent the 1$\sigma$ uncertainties in the last digit.

 For the compact corona models (I--III), the emission is dominated by the disk component, which is an extra problem for Model II where  the luminosity was near the Eddington limit. For the extended corona models (IV--V), the Comptonized emission dominates. 
  Models I--III yield  extremely low inner disk temperatures ($\sim$0.1--0.2 keV),  indicating strong retrograde BH spin for a $\sim$30 $M_{\rm BH}$ accretor. Such negative spin is unlikely, as only recently has there been observational evidence for a small minority of retrograd spins \citep[see e.g.][and references within]{steiner13, morningstar14,middleton14}; we note that \citet{middleton14} studied a XMM-Newton observation of a M31 ULX (PI R. Barnard); analysis of this spectrum revealed that the while the spin is negative for a 10 $M_\odot$ BH, it is positive for a 17 $M_\odot$ BH (their soft limit to the maximum BH mass). 

 Model III (compact corona and disk emission scattered by free electrons in the wind) yielded the best $\chi^2$/dof (383/422). However, we do not favor this model for two reasons. Firstly, the temperature ($\sim$0.2 keV) is very low for a $\sim$30 $M_\odot$ BH, as discussed above. Secondly, the thermal emission is absent from the scattered component, while the Comptonized emission is stronger for the scattered component than the direct component; this is more likely to represent  the high energy excess in the dip spectrum than a real scattering component.

 We  conclude that the corona is most likely to be  extended. 
If we set the absorption of the two emission components to be the same (Model IV), the best fit absorption = 6.4$\pm$1.7$\times 10^{24}$ atom cm$^{-2}$, with $\chi^2$/dof = 468/428. $\Delta\chi^2$ = 56 with respect to Model V, a $>$6$\sigma$ significance for 7 degrees of freedom. {\em Hence it is clear that the two emission components suffer different amounts of absorption, meaning that the dip evolution is most probably due to changes in the covering fraction rather than a simple change in opacity.}


We note that the background contributes $\sim$4\% of the non-dip spectrum and $\sim$20\% of the dip spectrum. However, our results are not particularly sensitive to the background; removing the background completely yields $\chi^2$/dof =437/428 for Model V, with $<$1$\sigma$ differences in parameter values.

\subsection{Estimating the size of the corona in IC10 X-1}

We estimated the size of the corona from the ingress, assuming that the absorber was located in the outer accretion disk. To do this, we estimated the disk radius to be 0.8$r_{\rm LBH}$ for the BH \citep[following][]{church04}, i.e. 0.30--0.34$a$.  From Equation 1, we estimate the corona size to be $\sim$0.1$a$.

 From Kepler's third law, we estimate $a$ to be $\sim$1.3--1.5$\times 10^{12}$ cm for the range of BH and WR masses discussed by \citet{silverman08}, and a corona radius $\sim$10$^{11}$ cm, a factor $<$2 times bigger than observed in the Galactic dipping source 4U1624$-$49 at a 1--30 keV luminosity $\sim$10$^{38}$ erg s$^{-1}$ \citep{church04}. The best fit 0.3--10 keV luminosity for IC10 X-1 = 3.7$\pm$0.3$\times 10^{38}$ erg s$^{-1}$, $\sim$0.1$L_{\rm EDD}$; the best fit 1--30 keV luminosity is 1.47$\pm$0.06$\times 10^{38}$ erg s$^{-1}$,
 hence this corona estimate is consistent with the  empirical relation derived by \citet{church04}. 

\begin{figure}
\epsscale{1.1}
\plotone{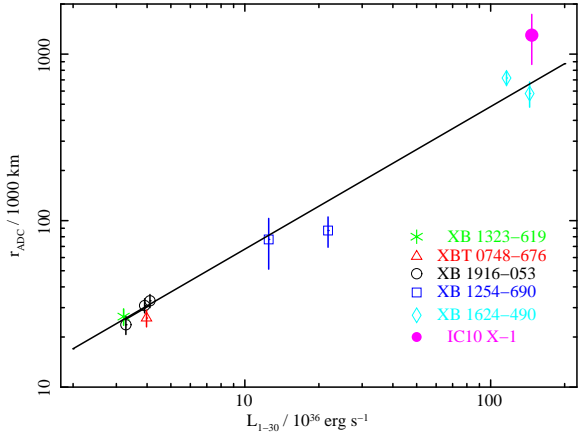}
\caption{Estimated corona size ($r_{\rm adc}$) vs. 1--30 keV luminosity for IC10~X-1, as well as 5 Galactic dipping NS XBs: XB~1323$-$619, XBT~0748$-$676, XB~1916$-$053, XB~1254$-$690, and XB~1624$-$490. Some XBs contribute data from multiple observations. The line represents the best fit to the NS XBs; this line and the uncertainties in the NS corona radii were estimated by eye from Figure 2 of \citet{church04}. }\label{adcvl}
\end{figure}

In Figure~\ref{adcvl} we present the estimated corona radius vs. 1--30 keV luminosity for IC10 X-1 and the five Galactic dipping NS XBs studied by \citet{church04}. The luminosities and corona sizes were gleaned from that paper, but the uncertainties and best fit were judged by eye; \citet{church04} did not cite the uncertainties for each point explicitly, but stated that the uncertainties were in the range 10--30\%. We set the 1$\sigma$ uncertainty in $r_{\rm ADC}$ for IC10 X-1 to 30\%.


We note that the scale of the corona is strongly related to the scale of the absorber, which is proportional to its distance from the X-ray emission regions.  We assume that the absorbing material is located at the edge of the disk; if it were instead located further in, for example near the circularization radius, then the corona sizes of IC10 X-1 and the Galactic dipping sources would all be smaller by similar amounts. If the absorption is caused by the WR wind, then the corona could be substantially larger.

Galactic NS XBs can exhibit dipping due to material in the outer disk for inclinations $\ga$60$^\circ$ \citep{frank87}. Furthermore, the minimum inclination for eclipsing is $>$75--78$^\circ$ for WR masses 17--35 $M_\odot$. If we assume the best fit mass function obtained by \citet{silverman08}, 7.64 $M_\odot$, then the allowed mass range for the BH in IC10 X-1 is $\sim$24--41 $M_\odot$ asssuming disk absorption with  inclination 60--78$^{\circ}$, slightly higher than the 23--38$M_\odot$ range quoted by \citet{silverman08}.

\subsection{Comparison of the dip size with the donor}

\citet{strohmayer13} estimated the deepest dip to last 5.2 hr, with a 3.3 hr ingress and a 4.6 hr egress, although the ingress was observed at the start of the observation, when the background radiation was high. These times correspond to $\sim$15\%, 10\%, and 14\% of the orbital period, with the total dip lasting $\sim$40\% of the orbital cycle. If the absorber is related to the secondary (either an eclipse or due to stellar winds), then the maximum dip requires an absorber size $\sim a$, and the total dip needs $\sim$2.4$a$.  

\citet{silverman08} estimated the black hole mass for a range of WR masses (17, 25, and 35 $M_\odot$) and inclinations (65, 78, and 90$^\circ$), and we calculated  the Roche lobe radius for the WR ($r_{\rm LWR}$) for each scenario. We found that $r_{\rm LWR}$ $\sim$0.34--0.38$a$, hence the Roche lobe is somewhat smaller than the maximum dip (0.96$a$). The long egress would require a thick stellar wind out to $>$3$r_{\rm LWR}$.  

\citet{clark04} found that the WR optical properties were well described by $\sim$85000 K emission with a luminosity 10$^{6.05}$ $L_\odot$, although other solutions were also possible. If we assume that the WR is emitting as a blackbody, then the optically thick part of the star has a radius $\sim$0.6--0.75 $r_{\rm LWR}$. Therefore, the absence of eclipsing rules out inclinations $\ga$75--80$^\circ$; a more massive WR allows  a higher inclination, because $a$ increases with the mass of the system.  We note that it is possible for the X-ray photosphere of the WR in IC10 X-1 to be substantially larger than the optical photosphere (our spectra see absorption equivalent to  $\ga$2$\times 10^{25}$ H atom cm$^{-2}$ as energy-independent); a larger photosphere would impose tighter constraints on the maximum inclination.

The X-ray lightcurve of IC10 X-1 contrasts with that of M33 X-7, another high inclination BH XB \citep{pietsch06a}. In this case, the donor (36--49 $M_\odot$) is $\sim$4--6 times more massive than the BH for various inclinations. \citet{pietsch06a} folded the 0.5--5.0 keV lightcurves from several Chandra observations, revealing a flat-bottomed dip where the minimum lasted $\sim$0.16 orbital cycles. Using Equation 2 we calculated the Roche lobe radius for the M33 X-7 donor to be $\sim$0.5$a$; this is entirely consistent with a true eclipse if the donor fills, or nearly fills, its Roche lobe. The light curve also shows scrappy dipping before the eclipse that may be due to an accretion stream that flows through the L1 point; Cygnus X-1 exhibits similar behavior \citep{balucinska-church00}.


\subsection{Considerations for windy absorber scenarios}

The WR wind system in IC10 X-1 is expected to be highly complex, and detailed modeling is well beyond the scope of this paper. \citet{clark04} found the observed properties of the WR to be well described by a model where the mass loss rate   is 4$\times 10^{-6}$ $M_\odot$ yr$^{-1}$. This value is considerably higher than the accretion rate required to power the observed  X-ray luminosity of $\sim4\times 10^{38}$ erg s$^{-1}$; if $L$ = $\eta \dot{M} c^2$, $\dot{M}$ is the accretion rate, and $\eta$ is the efficiency (assumed to be 0.1), then $\dot{M}$ $\sim$7$\times10^{-8}$ $M_\odot$ yr$^{-1}$. Only $\sim$2\% of the WR wind is required to power the X-ray source. 

Furthermore, the intense X-ray source is expected to substantially alter the structure of the wind. A large  portion of the wind will be highly ionized; if the ionization parameter, $\xi$, is calculated from $\xi = L_{\rm X}/nd^2$ where $n$ is the wind density, and $d$ is the distance from the X-ray source \citep{tarter69}, then $\xi$ $\sim$1000 at the location of  the WR star for reasonable parameters. However, the observed optical He line used by \citet{prestwich07} and \citet{silverman08} to calculate the mass function shows that some of the  gas is neutral; perhaps this neutral material is in the X-ray shadow of the WR. The ionization state of the absorbing material will greatly impact its absorbing signature.

The X-ray ionization also impacts the wind velocity, because the winds are line-driven \citep[see e.g.][]{castor75}. This complicates matters further, since the wind density (therefore ionization state) is affected by its velocity  \citep{hatchett77}. The fact that the WR is in a close binary system with a Roche lobe radius not much larger than WR radius complicates things yet further; for instance, it is not even clear whether the Galactic WR X-ray binary Cygnus X-3 contains a black hole or a neutron star, because much of the line emission is dominated or contaminated  by the wind  \citep[see e.g.][]{hanson00}.

One thing we can do is examine  the  absorption in the non-dip and dip spectra, and compare the difference with variations observed in other windy systems. For our favored model (Model V, assuming metalicity 0.15 Solar), $N_{\rm H}$ = 3.91$\pm$0.16$\times 10^{22}$ atom cm$^{-2}$ for the non-dip spectrum,  a factor $\sim$8 higher than the Galactic H{\sc i} column density in the direction of IC10 in  both the \citet{dickey90} survey and the Leiden/Argentine/Bonn (LAB) Survey \citep{kalberla05}. The absorption during deepest dipping is 8.4$\pm$1.4$\times 10^{23}$ atom cm$^{-2}$. Hence the absorption varies by a factor $\sim$20 over the orbital cycle. 

 We note that this is the equivalent column density for neutral hydrogen; the WR wind in IC10 X-1 has little hydrogen, and is unlikely to be neutral. Therefore the density of the ionized material needs to be substantially higher than the $N_{\rm H}$ equivalent to produce similar absorption. Such an ionized absorber is expected to produce absorption features similar to those seen in Galactic high inclination XBs \citep{boirin05,diaztrigo06}. However, any such features in IC10 X-1 are beyond the detection powers of current X-ray telescopes. 

\subsubsection{Comparison with the Galactic WR+BH binary Cygnus X-3}

Cygnus X-3 is the only Galactic WR + compact object binary known \citep{vankerkwijk96}, although \citet{mason12} claim that OAO 1657$-$415 either has a WR donor already, or soon will have. Even today the nature of the accretor  in Cygnus X-3 is unconfirmed due to complications caused by the wind; the evidence suggests a BH accretor, but is still consistent with a NS \citep{zdziarski13}. Cygnus X-3 exhibits a near sinusoidal X-ray lightcurve with a  4.8 hr period \citep{davidsen74} that is considerably shorter than that of IC10 X-1; it is also shorter than the period of NGC300 X-1, the only other known BH+WR XB \citep[$\sim$33 hr][]{carpano07}. 
The absorption experienced by Cygnus X-3 varies between $\sim$3$\times 10^{22}$ and 2$\times 10^{23}$ H atom cm$^{-2}$, assuming cold material with normal cosmic abundances; the absorbing material appears to be symmetrical around the donor star, i.e. asymmetrical with respect to the X-ray source \citep{parsignault72}. 

 \citet{zdziarski10} proposed that the X-ray variation of Cygnus X-3 included Compton down-scattering of the emission by a Thompson-thick, low temperature plasma cloud. \citet{zdziarski10} noted that scenarios  such as a wind-fed circumbinary envelope, or outflows from a disk are expected to be symmetrical around the accretor, and would not result in the observed flux variation; instead they suggest that the absorption may come from an inflated bulge where the stellar wind collides with the accretion disk.    They were unable to perform formal fits, due to the complexity of the model, but were able to reproduce observed spectra. With additional information from the power spectra, \citet{zdziarski10} were able to estimate the size of the cloud to be $\sim2\times 10^9$ cm, with a temperature $kT$ $\sim$3 keV and an optical depth $\sim$7.

The behavior exhibited by IC10 X-1 is strikingly different to the proposed behavior of Cygnus X-3, in that the high energy photons contribute a larger portion of the total flux in the dip spectrum than in the non-dip spectrum. At high energies, the  spectra are visibly converging in Fig.~\ref{fold}; also the models with energy-independent absorption systematically underestimated the high energy flux. An obvious difference between the two systems is in the orbital periods: 4.8 hr vs. $\sim$34 hr; accordingly, the accretor in Cygnus X-3 is likely to be embedded in a considerably more dense wind than the BH in IC10 X-1. We note that if we assume cosmic abundances for Model V, the non-dip absorption is just 7$\times 10^{21}$ atom cm$^{-2}$, only 2$\times 10^{21}$ cm$^{-2}$ above the line-of-sight absorption \citep{dickey90,kalberla05}, and a factor $\sim$15 lower than for Cygnus X-3.


\subsubsection{Comparison with the ultra-luminous X-ray source NGC5408 X-1}

Extra-galactic X-ray sources are often described as ``ultra-luminous'' if their luminosities are thought to be too high for stellar mass black holes, and are unassociated with the galaxy nucleus. Some  of these ultra-luminous X-ray sources (ULXs)  could be stellar mass black hole binaries in an special ultra-luminous accretion state \citep{gladstone09}, while others may contain intermediate mass black holes \citep[HLX1 in ESO 243-49 is a good candidate, see][]{farrell09}.

 \citet{strohmayer09} analyzed twice-weekly Swift observations of the well-known ULX NGC5408 X-1 over a 485 day interval, and found evidence for orbital modulation on a 115$\pm$4 day period.  Their spectral modeling suggested 0.3--10 keV luminosities up to $\sim$2$\times 10^{40}$ erg s$^{-1}$. The modulation was energy-dependent with an amplitude 0.24$\pm$0.02 below 1 keV and 0.13$\pm$0.02 above 1 keV; they suggest that the system has a 115 day orbital period, with a giant or supergiant donor. \citet{strohmayer09} considered the possibility that a super-orbital period was observed, but found it unlikey because it was shorter than any such period observed in BH XBs to date. They calculated the mean density for a Roche lobe-filling secondary to be 1.5$\times 10^{-5}$ g cm$^{-3}$, consistent with a giant or supergiant $\sim$5 $M_{\odot}$ donor and a 1000  $M_\odot$ accretor. 

However, this 115 day period was considerably weakened when the Swift observations were extended to $\sim$1200 days; instead, sharp intensity dips were observed on a $\sim$245 day period \citep{pasham13b,grise13}. \citet{pasham13b} studied the phase distribution of the 27 observed  dips assuming a 243 day period, which was consistent with a Gaussian with FWHM 0.24 in phase. The Galactic BH HMXB Cygnus X-1 exhibits short, energy-dependent dips that are concentrated at phase $\sim$0.7 that is thought to be due to the stream of material passing through the L1 point \citep{balucinska-church00}, and \citet{pasham13b} suggest that NGC5408 X-1 could be exhibiting similar behavior, and infer a high inclination ($\sim$70$^\circ$), similar to that of  IC10 X-1. However, we note that \citet{pasham13b} found no evidence for energy dependent dips in the Swift lightcurve of NGC5408 X-1. Furthermore, \citet{pasham13} analyzed non-dip and dip spectra, and the differences are consistent with a change in normalization rather than in column density.  

\citet{pasham13b} favor an orbital period $\sim$245 days, but note that it could be super-orbital. However, \citep{grise13} consider the $\sim$245 day period to be super-orbital, as they find that the strongest peak in their power spectrum is at $\sim$2.6 days, with other peaks at 115 and 183 days. We note that the M31 globular cluster XB Bo 158 (XB158) exhibits deep dipping on a $\sim$10 ks period  in some observations but not others, demonstrating that the disk is precessing \citep{barnard06}; we found that the luminosity of Bo 158 varied over $\sim$4--40$\times 10^{37}$ erg s$^{-1}$ during our $\sim$monthly Chandra monitoring program, and suggested that this luminosity variation was due to changes in accretion rate over the disk precession cycle \citep{barnard2012c}. It is therefore possible that the dips observed in NGC5408 X-1 are due to variations in accretion rate on the superorbital cycle also.

A shorter orbital period would suggest a lower BH mass than the 1000 $M_\odot$ assumed by \citet{strohmayer09} and \citet{pasham13b}. If NGC5408 X-1 exhibited a coplanar, extended corona as suggested by our modeling of IC10 X-1, then the observed  luminosity of $\sim 10^{40}$ erg s$^{-1}$ could be produced by the corona while keeping  the emission locally sub-Eddington. However, we note that there is some evidence for a 1000 $M_\odot$ BH from the energy spectrum and quasi-periodic oscillations exhibited by NGC5408 X-1 \citep[see][and references within]{strohmayer09}.

\subsubsection{The effect of absorber geometry on BH mass}

Another aspect to consider is whether the wind is spherical or aspherical. Knowledge of the wind geometry will constrain the inclination of the system, and therefore the BH mass. When calculating the BH mass, we adopt the mass function obtained by \citet{silverman08}, and estimate the BH mass for the range of WR masses used in that work. 

 The duration of the observed X-ray dipping suggests that the absorber  have a radius $\sim$3 $r_{\rm LWR}$, yet our spectral modeling suggests that $\sim$10--15\% of the emission is completely unabsorbed in the deepest dipping. If the absorber were spherical, this would require  either a low inclination ($\sim$40$^{\circ}$) or electron scattering in the wind. Since the non-dip and dip spectra converge at high energies (see Fig. 3), the residual emission is unlikely to be dominated by scattering. 

Such a low inclination would suggest a BH mass $\sim$50--65 $M_\odot$ for a WR with mass 17--35 $M_\odot$.
A 65 $M_\odot$ BH is not impossible, if it was formed via direct collapse of a metal poor, high mass star so that little mass was lost during the main sequence lifetime \citep{belczynski10}. We note that \citet{belczynski10} cite IC10 X-1 as an example of a star born in a galaxy with moderate metalicity (they assume 0.3 Solar), with a BH mass that is in good agreement with what they expect for such metalicities. \citet{belczynski10} found the maximum BH mass to be $\sim$15 $M_\odot$ for Solar metalicity, $\sim$30 $M_\odot$ for 30\% Solar, and $\sim$80 $M_\odot$ for 1\% Solar. We are unable to simply interpolate the maximum mass for 15\% Solar metalicity, but it seems unlikely to be as high as 50--65 $M_\odot$.

One possible cause for an aspherical WR wind is stellar rotation. \citet{harries98} examined a population of single WR stars, and found evidence  that 20\% of the sample have flattened winds, using linear spectropolarimetry. Also, 
\citet{hanson00} suggested that the wind of the donor in Cygnus X-3 is highly perturbed, and may be consistent with the presence of a disk wind; this  may be in the orbital plane. 

Another possible cause for asymmetrical absorption in the wind would be the gas stream \citep[as seen at phase 0.6 in Cygnus X-1, ][]{balucinska-church00}, so long as the inclination is high enough. \citet{blondin91} performed two dimensional simulations of such streams, finding that density of the stream is enhanced by a factor 20--30; this is similar to the observed density variation in IC10 X-1. Furthermore, the tidal stream would produce stable absorption features that are fixed in phase, and that last for a substantial portion of the orbital cycle, again similar to IC10 X-1. While this scenario is appealing, there are two caveats. Firstly, \citet{blondin91} assumed an inclination of 90$^\circ$ for their simulations; the stream would likely have a reduced effect at lower inclinations. Secondly, the strength of the gas stream was found to be strongly dependent on the separation, $a$,  between the stars; the simulations covered separations of 1.5--1.7 stellar radii, while in IC10 X-1, the separation is 4--5 WR radii. 

In order to avoid energy-independent eclipsing by the WR, and yet still see a aubstantial stream,  the inclination would likely be  $\sim$75--78$^\circ$. Using the \citet{silverman08} mass function and WR mass estimates yields a BH mass $\sim$24--33 $M_\odot$ for this scenario.

 \section{Discussion and Conclusions}

We have analyzed the spectral and temporal evolution of the BH+WR XB IC10 X-1, during a 130 ks XMM-Newton observation. IC10 X-1 exhibits X-ray intensity dips on a $\sim$35 hr orbital period, and we considered three scenarios: energy-independent eclipses, and energy-dependent absorption from either material in the accretion disk or the WR wind.

Analyzing the 0.3--1.0 keV and 4.0--10 keV lightcurves showed the probability for a constant  hardness ratio ($I_{\rm 4.0-10 keV}/I_{\rm 0.3-1.0 keV}$) was $\sim$2$\times 10^{-8}$, rejected at a 5.6$\sigma$ level. Therefore, the intensity dips cannot be due to simple occultation by the WR, and must come from a lower-density region such as the outer accretion disk or the WR wind. 
Our favoured results were obtained from a model consisting of a disk blackbody component that was completely removed during the dip, and a partially-covered Comptonized component; a similar  model has been successfully applied to high-inclination NS XBs in our Galaxy \citep[see e.g.][]{church95,smale01,barnard01,smale02}.

One model for such a system is the two-phase accretion disk proposed by \citet{haardt93}; in this scenario, a coplanar, hot corona embeds the cold disk, and is fed by cool photons from the disk while hot photons from the corona heat the disk.

If we assume that the dip is caused by material in the outer disk, then we may estimate the size of the corona from the dip ingress; the thermal and Comptonized emission regions in Galactic dipping sources increase in absorption at different rates during dip evolution, meaning that the ingress is a change in absorbed fraction rather than a change in density \citep[see][and references within]{church04}. We estimate the corona size to be $\sim$10$^6$ km. Remarkably,  this result  is consistent with the relation found between corona size and 1--30 keV luminosity in Galactic dipping NS XBs \citep{church04};  the 1--30 keV luminosity was 1.40$\pm$0.06$\times 10^{38}$ erg s$^{-1}$. For this  scenario the inclination is likely to be $\sim$60--80$^\circ$, giving a BH mass $\sim$24--41 $M_\odot$,  which is slightly larger than the current estimate.


If the absorption is due to the WR wind, then this wind is unlikely to be spherical. A spherical wind would require an inclination $\sim$40$^\circ$, and a BH mass $\sim$50--65 $M_\odot$; however, \citet{belczynski10} found that the maximum mass for a stellar mass BH formed in a moderately metal poor galaxy like IC10 to be $\sim$30 $M_\odot$ (although they assumed a metalicity $\sim$30\% Solar).
It is possible that the absorption is caused by the gas stream that results from part of the stellar wind flowing through the L1 point. For this scenario, the inclination would need to be as high as is allowed by the absence of energy-independent eclipsing, $\sim$75--78$^\circ$. The BH mass would therefore be $\sim$24--33 $M_\odot$, within the current range. We note that an absorbing  wind would need to be considerably larger than an absorber in the disk to produce the same ingress and egress times, hence the corona would be similarly larger.

 While it is unclear whether the absorbing material originates in the accretion disk or WR wind, our results provide hard evidence  that IC10 X-1 has a  substantial extended corona. It would be  capable of producing a total luminosity that exceeds the Eddington luminosity while remaining locally sub-Eddington. Therefore, many ULXs could  do this too. This would allow stellar mass BHs to power many of the ULXs observed to date.

\section*{Acknowledgments}
We thank the referee for their insights  on how to improve the paper. We thank Paul Crowther for useful discussions about the properties of the wind.  This research has made use of  data from XMM-Newton, an ESA science mission with instruments and and contributions directly funded by ESA member states and the US (NASA).  RB was supported by NASA grant GO3-14095X. JFS was supported by NASA Hubble Fellowship grant HST-HF-51315.01.

{\it Facilities:} \facility{XMM-Newton (pn)}






\clearpage



\end{document}